\documentclass[prd,showpacs,amsmath,amssymb,nofootinbib]{revtex4}
\usepackage{graphicx,color}
\begin{document}
\title{$\Lambda_{c}(2940)^{+}$: A Possible Molecular State?}
\vspace{0.8cm}

\author{Xiao-Gang He$^{1,2}$, Xue-Qian Li$^1$, Xiang Liu$ ^1$ and Xiao-Qiang Zeng$^1$}

\vspace{1cm}

\affiliation{1. Department of Physics, Nankai University, Tianjin
300071.\\
2. Department of Physics and Center for Theoretical Sciences,
National Taiwan University, Taipei, 1067}
\date{\today}

\vspace{2cm}

\begin{abstract}

A new baryonic state $\Lambda_{c}(2940)^{+}$ has recently been
discovered by the Babar collaboration in the $D^{0}p$ channel.
Later Belle collaboration also observed this state in the
$\Sigma_c(2455)^{0,++} \pi^{\pm}\to \Lambda_{c}^{+}\pi^{+}\pi^{-}$
channel. The mass of $\Lambda_{c}(2940)^{+}$ is just a few MeV
below the sum of $D^{*0}$ and $p$ masses suggesting a possibility
that this state may be a $D^{*0} p$ molecular state. In this paper
we study whether such a molecular state can be consistent with
data. We find that the molecular structure can explain data and
that if $\Lambda_c(2940)^+$ is a $D^{*0} p$ molecular state it is
likely a $1/2^-$ state. Several other decays modes are also
suggested to further test the molecular structure of
$\Lambda_c^+(2940)$.
\end{abstract}
 \maketitle

\section{Introduction}

Very recently, the Babar collaboration announced that a new charmed
baryonic state $\Lambda_{c}(2940)^{+}$ has been observed in the mass
spectrum of $D^0 p$ \cite{babar}. Its mass and width are
respectively
$$m=2939.8\pm 1.3(\mathrm{stat.})\pm 1.0(\mathrm{syst.})\;
\mathrm{MeV}/\mathrm{c}^2$$ and $$\Gamma=17.5 \pm
5.2(\mathrm{stat.})\pm5.9(\mathrm{syst.})\; \mathrm{MeV}.$$ Its
spin and parity have not been determined by experimental
measurement yet. Another charmed baryonic state
$\Lambda_{c}(2880)^+ $ is also observed in the $D^0 p$ spectrum by
the Babar collaboration. The state $\Lambda_{c}(2880)^+ $ had
already been observed before by the CLEO collaboration in the mass
spectrum of $\Lambda_{c}^{+}\pi^{+}\pi^{-}$ \cite{CLEO} and the
Belle collaboration has recently also observed
$\Lambda_c(2940)^{+}$ in $\Lambda_c^+ \pi^+\pi^-$ channel via
$\Lambda_c(2940)^+\to \Sigma_c(2455)^{0,++} \pi^{\pm} \to
\Lambda_c^+ \pi^+\pi^-$\cite{belle}.

\begin{figure}[htb]
\begin{center}
\begin{tabular}{cc}
\scalebox{0.6}{\includegraphics{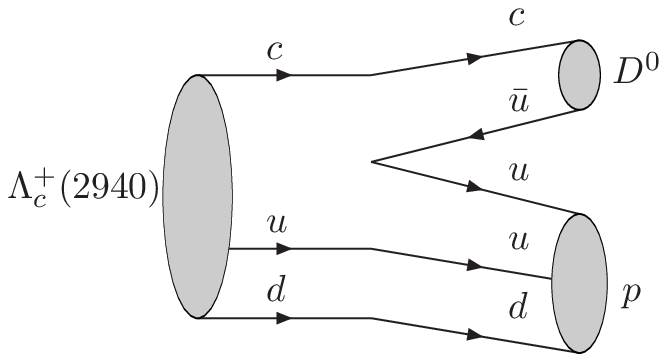}}&\scalebox{0.6}{\includegraphics{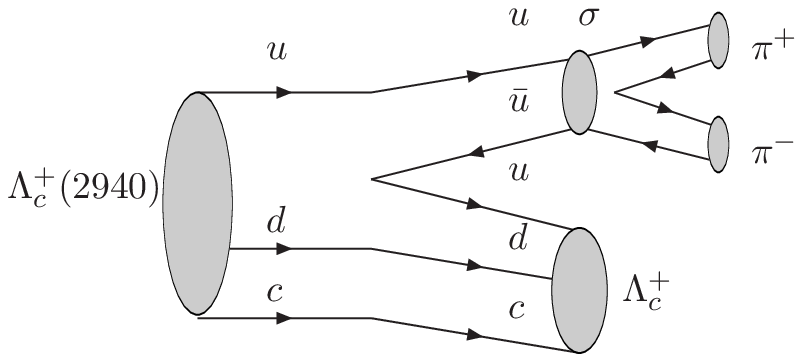}}
\\(a)&(b)\\
\end{tabular}
\end{center}
\caption{Decays of excited charmed baryon $\Lambda_{c}^{+}$ to (a)
 $D^{0}p$,  and
 (b) $\Lambda_{c}^{+}\pi^{+}\pi^{-}$.} \label{diagram}
\end{figure}

As commonly believed, $\Lambda_{c}(2880)^+ $ can be categorized as
an excited charmed baryon \cite{2880-1,2880-2,2880-3}. It has
large branching ratios into both $D^{0}p$ and
$\Lambda_{c}^{+}\pi^{+}\pi^{-}$ decay modes which can be realized
via a subsequent process as shown in Fig\ref{diagram}. The new
$\Lambda_c(2940)^+$ might also be an excited state, but the sum of
the masses of $D^{*0}$ and $p$, ($m_{D^{*0}}+m_{p}=2945$ MeV), is
so close to the required 2940 MeV makes it very tempting to view
it as a $D^{*0} p$ molecular state. The slight excess energy above
the central value of $\Lambda_{c}(2940)^{+}$ mass can be
attributed to the binding energy of the two constituents. If
indeed $\Lambda_c(2940)$ is a molecular state, the decay pattern
may be different. In the following we study if a $D^{*0} p$
molecular state is consistent with data.

\section{$D^{*0} p$ Molecular states}

There is an abundant spectrum in charm-tau energy range, some of
the charmed states are very close to each other in masses, their
peaks even overlap. Some of them may be molecular states. The
picture of molecular states was first proposed to interpret the
behaviors of scalar mesons $f_{0}(980)$ and $a_{0}(980)$ which
could be $K\bar{K}$ molecules \cite{980-1,980-2,980-3,Tuan}. This
idea has now been widely adopted for explaining some experimental
data. Rujula, Geogi, and Glashow suggested that $\psi(4040)$ is a
$D^{*}\bar{D}^{*}$ molecular state \cite{RGG}, Rosner and Tuan
also studied the so-called C-exotic states \cite{Tuan}. In a
recent work, it was suggested that $Y(4260)$ is a
$\chi_{c}-\rho^{0}$\cite{Y4260} or $\chi_{c}-\omega$ \cite{Yuan}
molecule. For the dynamics, Okun and Volosin studied the
interaction between charmed mesons and molecular states involving
charmed quarks \cite{Okun}. Following the previous studies it is
not unrealistic to consider $\Lambda_c(2940)^{+}$ as a molecular
state.

An immediate question one needs to answer, if interpreting
$\Lambda_{c}(2940)^{+}$ as a $D^{*0} p$ molecular state, is that
whether the correct binding energy of about 5 MeV can be realized.
We find that one particle exchange model can indeed achieve this.
In this model one deduces the effective potential of $D^{*0}p$
system by using the linear $\sigma$ model \cite{Liu}. Following
the standard procedures \cite{Landau}, we calculated the
transition matrix element of the elastic scattering
$D^{*0}+p\rightarrow D^{*0}+p$ in the momentum space by regular
quantum field theory method, and then setting $q_0=0$ where $q^0$
is the 0-th component of the momentum of the exchanged hadrons
which possess appropriate quantum numbers, a Fourier
transformation from the momentum space to the configuration space
is then carried out to obtain the potential. This is the effective
potential between the constituents $D^{*0}$ and $p$. Substituting
this effective potential into the Schr\"{o}dinger equation, one
can obtain the wave function and eigen-energy which is identified
to be the binding energy of the molecular state.

We have carried out an calculation similar to that in \cite{Liu}
by taking exchanges of $\pi$, $\omega$ and $\rho$ mesons as the
leading interaction of $D^{*0}$ and $p$ which eventually binds
$D^{*0}$ and $p$ into a molecular state, to obtain the binding
energy. In the derivation of the effective interaction in the
momentum space, to compensate the off-shell effects for the
exchanging mesons, one usually phenomenologically introduces a
form factor at each effective vertex. A commonly used form factor
is taken as \cite{formfactor}
$(\Lambda^{2}-M_{m}^{2})/(\Lambda^{2}-q^2)$, which we also use in
this work. Here $\Lambda$ is a phenomenological parameter whose
value is near 1 GeV \cite{footnote}, and $q$ is the four-momentum
of the exchanged meson.

Since the spin of $\Lambda_{c}(2940)^{+}$ is not determined by
experiment yet, we consider two possible cases for the spin of
$\Lambda_{c}(2940)^{+}$ in the S-wave state. In this case one
obtains two $J^P$ states: (a) $1/2^-$, and (b) $3/2^-$. We display
the allowed ranges for the masses of the molecular states and
their binding energies in Table \ref{haha}. We see that the
binding energies are in the right ranges. From the spectrum we
cannot distinguish whether the spin is 1/2 or 3/2. Adjusting
parameters in the form factors, P-wave states with the right
binding energy is also possible. In that case one would have
$1/2^+$ and $3/2^+$ states. To get more information, one needs to
invoke the decay rates measured recently by the Babar and Belle
collaborations.
\\

\begin{table}[htb]
\begin{center}
\begin{tabular}{|c|c|} \hline
For case (a) $1/2^-$ &For case (b) $3/2^-$ \\\hline
$\Lambda=0.85\sim0.89$ GeV &$\Lambda=0.90\sim0.95$ GeV\\\hline
$E_{D^{*0}p}=-7.2\sim-2.9$ MeV&$E_{D^{*0}p}=-7.4\sim-3.6$
MeV\\\hline $m_{_{D^{*0}p}}=2.938\sim2.942$
GeV&$m_{_{D^{*0}p}}=2.938\sim2.941$ GeV\\\hline
\end{tabular}
\end{center}
\caption{The binding energies and the masses correspond to S-wave
$D^{*0}p$ systems with spin 1/2 and spin 3/2
respectively.}\label{haha}
\end{table}

\section{$\Lambda_c(2940)^+\to D^0 p, \Lambda_c^+ \pi^+\pi^-$}

\begin{figure}[htb]
\begin{center}
\begin{tabular}{cc}
\scalebox{0.9}{\includegraphics{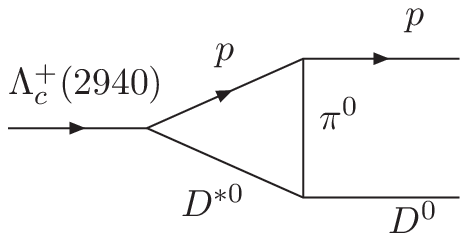}}
&\scalebox{0.9}{\includegraphics{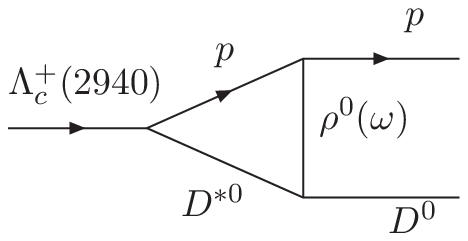}}
\\(a)&(b)\\
\end{tabular}
\end{center}
\caption{} \label{diagram-2}
\end{figure}

\begin{figure}[htb]
\begin{center}
\begin{tabular}{cc}
\scalebox{0.9}{\includegraphics{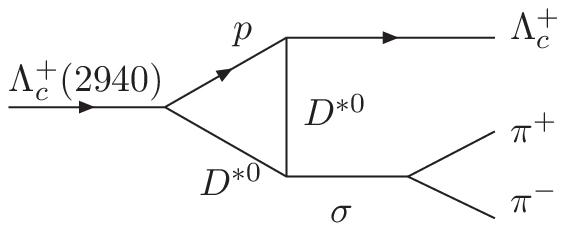}}&\scalebox{0.9}{\includegraphics{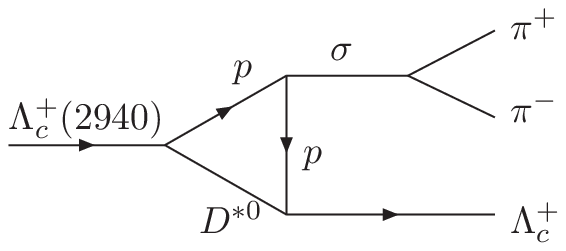}}
\\(a)&(b)\\
\end{tabular}
\end{center}
\caption{} \label{diagram-3}
\end{figure}

\begin{figure}[htb]
\begin{center}
\begin{tabular}{ccc}
\scalebox{0.9}{\includegraphics{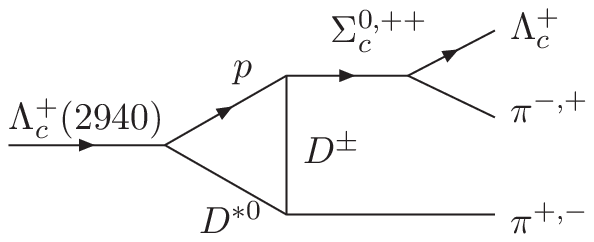}}&\scalebox{0.9}{\includegraphics{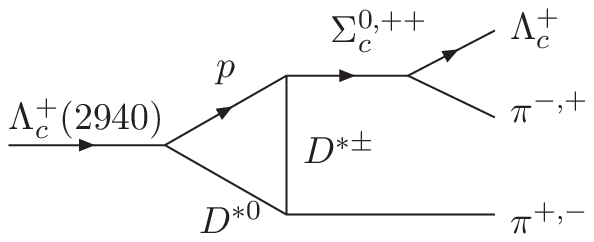}}
&\scalebox{0.9}{\includegraphics{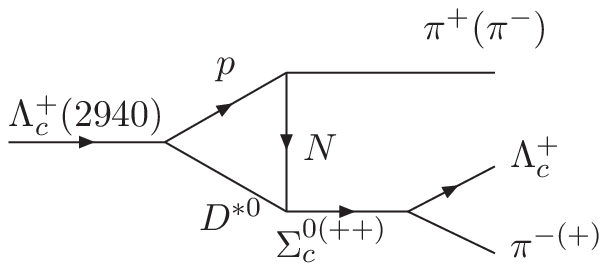}}
\\(a)&(b)&(c)\\
\end{tabular}
\end{center}
\caption{} \label{diagram-4}
\end{figure}

To have more information about the $D^{*0} p$ molecular state, we
now consider the decay of this state to $D^0 p$ and $\Lambda^+_c
\pi^+ \pi^-$. Since $\Lambda_c(2940)^+$ is just below the threshold
of $D^{*0} p$, it can fall apart into $D^{*0} p$ through threshold
effects due to finite width. More specifically, we assume that the
dominant decays of $\Lambda_c(2940)^{+}$ occur via two steps shown
in Fig. \ref{diagram-2}(\ref{diagram-3},\ref{diagram-4}). The
molecular state $\Lambda_{c}(2940)^{+}$ first dissolves into
$D^{*0}$ and $p$ due to the threshold effect, that is, the finite
width of $\Lambda_c^+(2940)$ about 20 MeV allows on-shell final
$D^*$ and $p$ with small three momenta. Thus $D^{*0}$ and $p$ are
treated as on-mass-shell real particles, and then $D^{*0}$ and $p$
re-scatter into $D^{0}p$ or $\Lambda_{c}^{+}\pi^+\pi^-$ by
exchanging intermediate states. If $\Lambda_c(2940)^+$ is a
${1/2}^{+}$ or ${3/2}^{+}$ state, it must be a P-wave molecule of
$D^*$ and $p$ and its dissociation into $D^* p$ (or derivative of
the wave function $\Psi'(0)$) is further suppressed by small three
momentum. Thus the transition rate would be very small. In this
picture, $\Lambda_c(2940)^+$ is disfavored to be P-wave or higher
wave bound states. We then left with S-wave $1/2^-$ and $3/2^-$ to
study.

We now proceed to calculate $\Lambda_c(2940)^+\to D^0 p$ and
$\Lambda_c(2940)^+ \to \Lambda_c^+ \pi^+ \pi^-$ decays by the
diagrams in Fig. \ref{diagram-2}(\ref{diagram-3},\ref{diagram-4})
with on-shell $D^{*0}$ and $p$, and then $D^{*0}$ and $p$ re-scatter
in to the desired final states. For the $\Lambda_c(2940)^+$ coupling
to $D^{*0}$ and $p$, we write as\cite{chung}
\begin{eqnarray}
L_{bound} =
g_{_{{ND^{*0}\Lambda_{c}^{+}(2940)}}}\bar{N}A_{\mu}\epsilon^{\mu}_{D^{*0}},
\end{eqnarray}
and
\begin{eqnarray*}
A_{\mu}=\left\{\begin{array}{ll}\gamma_{\mu}\gamma^{5}\Lambda_{c}(2940)^{+},&J^{P}(2940)=\frac{1}{2}^{-},
\\{\Lambda_{c}}_{\mu}^{+}(2940),&J^{P}(2940)=\frac{3}{2}^{-},
\end{array}\right .
\end{eqnarray*}
where ${\Lambda_{c}}_{\mu}^{+}$ is the Rarita-Schwinger
vector-spinor for a spin-3/2 particle. For $\Lambda_c^+(2940)$ to
be $1/2^-$ and $3/2^-$, $A_\mu$ are given by:
$\gamma_{\mu}\Lambda_{c}^{+}(2940)$ and
$\gamma^{5}{\Lambda_{c}}_{\mu}^{+}(2940)$, respectively. In the
above $g_{{ND^{*}\Lambda_{c}^{+}(2940)}}$ parameterizes the bound
state effect which is not known. Since we will be concerned with
relative strength of $\Lambda_c^+(2940)\to D^0 p$ and
$\Lambda_c^+(2940)\to \Lambda_c^+ \pi^+\pi^-$, the specific value
of $g_{{ND^{*}\Lambda_{c}^{+}(2940)}}$ is not important for our
purpose.

For the couplings of the other vertices in  Fig.
\ref{diagram-2}(\ref{diagram-3},\ref{diagram-4}), we write the
relevant Lagrangian as
\begin{eqnarray}\label{lagrangian}
L&=&g_{{NN\pi}}\bar{\psi}(\sigma+i\gamma_{5}\mathbf{\tau\cdot\pi})\psi
+g_{{NN\rho}}\bar{\psi}\gamma_{\mu}\mathbf{\tau}\psi\cdot\mathbf{\rho}^{\mu}
+g_{{VV\pi}}\varepsilon^{\mu\nu\alpha\beta}
\partial_{\mu}V_{\nu}^{\dag}\mathbf{\tau}\partial_{\alpha}V_{\beta}\cdot\mathbf{\pi}\nonumber\\
&&
+g_{{VV\sigma}}[\partial^{\mu}V^{\dag\nu}\partial_{\mu}V_{\nu}-\partial^{\mu}
V^{\dag\nu}\partial_{\nu}V_{\mu}]\sigma+
g_{{VV\rho}}[(\partial_{\mu}V^{\dag\nu}\mathbf{\tau}V_{\nu}-\partial^{\nu}
V^{\dag}_{\mu}\mathbf{\tau}V_{\nu})\cdot\rho^{\mu}\nonumber\\
&&+(V^{\dag\nu}\mathbf{\tau}\cdot\partial_{\mu}\rho_{\nu}-\partial_{\mu}
V^{\dag\nu}\mathbf{\tau}\cdot\rho)V^{\mu}+
V^{\dag\mu}(\tau\cdot\rho\partial_{\mu}V_{\nu}-\tau\cdot\partial_{\mu}\rho
V_{\nu})]\nonumber\\
&&
+[g_{{VP\pi}}V^{\dag\mu}\mathbf{\tau}\cdot(P\partial_{\mu}\mathbf{\pi}-\partial_{\mu}P\pi)+h.c.]
+g_{{VP\rho}}\varepsilon^{\mu\nu\alpha\beta}[\partial_{\mu}\mathbf{\rho}_{\nu}\partial_{\alpha}
V^{\dag}_{\beta}\cdot\mathbf{\tau}P
+\partial_{\mu}V^{\dag}_{\nu}\mathbf{\tau}\cdot\partial_{\alpha}\mathbf{\rho}_{\beta}P]\nonumber\\
&&
+g_{{ND^{*}\Lambda_{c}^{+}(2285)}}\bar{N}\gamma_{\mu}\Lambda_{c}^{+}(2285)D^{*\mu}
+g_{{\Sigma_{c}^{+}(2455)D^{*}N}}\bar{N}\gamma_{\mu}\Sigma_{c}^{+}(2455)D^{*\mu}
+g_{{\Sigma_{c}^{+}(2520)D^{*}N}}\bar{N}\gamma_{5}\Sigma_{c}^{+\mu}(2520)D^{*\mu}\nonumber\\
&&
+g_{{\Sigma_{c}^{+}(2455)DN}}\bar{N}\gamma_{5}\Sigma_{c}^{+}(2455)D
+\frac{g_{{\Sigma_{c}^{+}(2520)DN}}}{m_{_{D}}}\bar{\Sigma}_{c}^{+\mu}(2520)N(\partial_{\mu}D),
\end{eqnarray}
where P, V are pseudoscalar and  vector mesons $D, D^{*}$. The
known couplings are given by $g_{_{NN\pi}}=13.5$,
$g_{{NN\rho}}=3.25$, $g_{{D^{*}D^{*}\rho}}=2.9$,
$g_{{D^{*}D\pi}}=18$, $g_{{D^{*}D\rho}}=4.71$ GeV$^{-1}$ and
$g_{{D^{*}D^{*}\pi}}=g_{{D^{*}D^{*}\sigma}}=3.5$ \cite{Liu,cheng}.
The values of the couplings $g_{{N D^{*}\Lambda_{c}^{+}(2285)}}$,
$ g_{{N D\Sigma_{c}(2455)}}$, $ g_{{N D^{*}\Sigma_{c}(2455)}}$, $
g_{{ND\Sigma_{c}(2520) }}$ and $ g_{{ND^{*}\Sigma_{c}(2520) }}$
are not known. We keep them here and discuss their effects later.

From the above effective Lagrangian we can construct decay
amplitudes for $\Lambda_c^+(2940) \to D^0 p, \Lambda_c^+ \pi^+\pi^-$
through the triangle diagrams shown in  Fig.
\ref{diagram-2}(\ref{diagram-3},\ref{diagram-4}). These amplitudes
are given in the Appendix.

Once the decay amplitudes are obtained, the partial decay width
can be derived. We give three typical examples in the following
\begin{eqnarray}
&&\Gamma(\Lambda_{c}(2940)^{+}\to D^{0}p)=
\frac{1}{2M}\int\frac{d^{3}P{_{D^{0}}}}{(2\pi)^3}\frac{1}{2
E_{D^{0}}}\frac{d^{3}P{_{p}}}{(2\pi)^3}\frac{2m_{p}}{2
E_{p}}(2\pi)^4
\delta^4(M-P{_{D^{0}}}-P{_{p}})\nonumber \\
&&\hspace{3.2cm}\times|\mathcal{M}(\Lambda_{c}(2940)^{+}\to D^{0} p)|^2,\nonumber\\
&&\Gamma(\Lambda_{c}(2940)^{+}\to \Lambda_c^+ \sigma \to
\Lambda_{c}^{+}\pi^{+}\pi^{-})_{\sigma}=
\frac{1}{2M}\int\frac{d^{3}P{_{\Lambda_{c}^{+}}}}{(2\pi)^3}\frac{2m_{\Lambda_{c}^{+}}}{2
E_{\Lambda_{c}^{+}}}\frac{d^{3}P{_{\sigma}}}{(2\pi)^3}\frac{1}{2
E_{\sigma}}(2\pi)^4
\delta^4(M-P{_{\Lambda_{c}^{+}}}-P{_{\sigma}})\nonumber \\
&&\hspace{3.2cm}\times|\mathcal{M}(\Lambda_{c}(2940)^{+}\to
\Lambda_{c}^{+} \sigma)|^2\times
B(\sigma\to\pi^{+}\pi^{-}),\nonumber\\
&&\Gamma(\Lambda_{c}(2940)^{+}\to\Sigma_c \pi \to
\Lambda_{c}^{+}\pi^{+}\pi^{-})_{\Sigma_{c}}=
\frac{1}{2M}\int\frac{d^{3}P{_{\Sigma_{c}}}}{(2\pi)^3}\frac{2m_{\Sigma_{c}}}{2
E_{\Sigma_{c}}}\frac{d^{3}P{_{\pi}}}{(2\pi)^3}\frac{1}{2
E_{\pi}}(2\pi)^4
\delta^4(M-P{_{\Sigma_{c}}}-P{_{\pi}})\nonumber \\
&&\hspace{3.2cm}\times|\mathcal{M}(\Lambda_{c}(2940)^{+}\to
\Sigma_{c} \pi)|^2\times
B(\Sigma_{c}\to\Lambda_{c}^{+}\pi).\nonumber
\end{eqnarray}
Numerically the branching ratio $B(\sigma\to\pi^{+}\pi^{-})$ is
about 0.6$\sim 0.7$ and the branching radio
$B(\Sigma_{c}\to\Lambda_{c}^{+}\pi)$ is about 1. \cite{PDG}. Note
that since $g_{{ND^{*}\Lambda_{c}^{+}(2940)}}$ is not known, we
will not be able to obtain the absolute value. However, we can
obtain the relative strength of each diagram which can still give
information about the decay pattern of $\Lambda_c^+(2940)$.

\section{Results and Discussions}

We now discuss contributions from each diagram to
$\Lambda^+_c(2940) \to D^0 p$ and $\Lambda^+_c(2940)\to
\Lambda^+_c \pi^+\pi^-$. Here we have omitted (and will omit) to
indicate the intermediate stage $D^*p$ in the decay chain. Since
the spin of $\Lambda_{c}(2940)^{+}$ has not been determined, we
discuss both $J^P = 1/2^-$ and $J^P=3/2^-$.

(1) $J^{P}(\Lambda_{c}^{+})=\frac{1}{2}^{-}$.

For $\mathcal{M}(\Lambda_{c}(2940)^{+}\to  D^{0}p)$ with $\pi$ and
$\rho^{0}$ exchange, we have
\begin{eqnarray*}
|\mathcal{M}(\Lambda_{c}(2940)^{+}\to
D^{0}p)_{\pi^{0}}|^2:|\mathcal{M}(\Lambda_{c}(2940)^{+}\to
D^{0}p)_{\rho^{0}}|^2=1.0\times10^3:1.
\end{eqnarray*}

Since the $\Lambda_c^+(2940)$ wave function is not known (in our
notation the coupling $g_{{ND^{*}\Lambda_{c}^{+}(2940)}}$ is not
known), it is not possible to calculate the absolute value for
each of the partial decay width. However, we are able to obtain
the relative widths which can also provide us with useful
information. For convenience of discussion, in the above we have
normalized $|\mathcal{M}(\Lambda_{c}(2940)^{+}\to
D^{0}p)_{\rho^{0}}|^2=1$. The matrix element squared listed below
are all evaluated according to the same normalization.

The contribution to the amplitude from $\omega$ exchange is close
to that from $\rho$ exchange, that is,
$\mathcal{M}(\Lambda_{c}(2940)^{+}\to D^{0}p)_{\rho^{0}}\simeq
\mathcal{M}(\Lambda_{c}(2940)^{+}\to D^{0}p)_{\omega}$. So when we
consider the decay of $\mathcal{M}(\Lambda_{c}(2940)^{+}\to
D^{0}p)$, we can almost ignore the influence of $\rho$ and
$\omega$ exchange and consider $\pi^{0}$ exchange only.

For $\mathcal{M}(\Lambda_{c}^{+}(2940)\to
\Lambda_{c}^{+}\pi^{+}\pi^{-})$, we have considered several
contributions, $\mathcal{M}(\Lambda_{c}(2940)^{+}\to
\Lambda_{c}^{+}\sigma\to\Lambda_{c}^{+}\pi^{+}\pi^{-})$,
$\Lambda_{c}^{+}(2940)\to
\Sigma_{c}(2455)\pi\to\Lambda_{c}^{+}\pi^{+}\pi^{-})$ and
$\Lambda_{c}^{+}(2940)\to
\Sigma_{c}(2520)\pi\to\Lambda_{c}^{+}\pi^{+}\pi^{-})$ with exchanges
of different intermediate states in various triangle diagrams. For
$\mathcal{M}(\Lambda_{c}(2940)^{+}\to
\Lambda_{c}^{+}\sigma\to\Lambda_{c}^{+}\pi^{+}\pi^{-})$, we have
\begin{eqnarray*}
|\mathcal{M}(\Lambda_{c}(2940)^{+}\to
\Lambda_{c}^{+}\sigma)_{D^{*}}|^2:|\mathcal{M}(\Lambda_{c}(2940)^{+}\to
\Lambda_{c}^{+}\sigma)_{N}|^2=2.0\times10^{-4}:0.015\;.
\end{eqnarray*}
In this case $N$ exchange is more important. The resulting branching
ratio for $\Lambda_{c}^+(2940) \to \Lambda_c^+ \pi^+\pi^-$ will be
much smaller than that for $\Lambda_{c}^+(2940)\to D^0 p$.

For $\mathcal{M}(\Lambda_{c}^{+}(2940)\to
\Sigma_{c}(2455)\pi\to\Lambda_{c}^{+}\pi^{+}\pi^{-})$, we have
\begin{eqnarray*}
&&|\mathcal{M}(\Lambda_{c}(2940)^{+}\to
\Sigma_{c}(2455)\pi)_{D}|^2:|\mathcal{M}(\Lambda_{c}(2940)^{+}\to
\Sigma_{c}(2455)\pi)_{D^{*}}|^2:
|\mathcal{M}(\Lambda_{c}(2940)^{+}\to
\Sigma_{c}(2455)\pi)_{N}|^2\nonumber\\
&&=0.02g_{{ND\Sigma_{c}^{+}(2455)}}^2:
1.5\times10^{-3}g_{{ND^{*}\Sigma_{c}^{+}(2455)}}^2:9.3g_{{ND^{*}\Sigma_{c}^{+}(2455)}}^2\;.
 \end{eqnarray*}
Since the couplings are not determined, the numbers cannot be
determined. If the couplings are similar in order of magnitude, the
$N$ exchange is most likely to dominate. We will come back to
discuss this later.

For $\mathcal{M}(\Lambda_{c}^{+}(2940)\to
\Sigma_{c}(2520)\pi\to\Lambda_{c}^{+}\pi^{+}\pi^{-})$, we have
\begin{eqnarray*}
&&|\mathcal{M}(\Lambda_{c}(2940)^{+}\to
\Sigma_{c}(2520)\pi)_{D}|^2:|\mathcal{M}(\Lambda_{c}(2940)^{+}\to
\Sigma_{c}(2520)\pi)_{D^{*0}}|^2:|\mathcal{M}(\Lambda_{c}(2940)^{+}\to
\Sigma_{c}(2520)\pi)_{N}|^2\nonumber\\
&&=4.2\times10^{-2}g_{{ND\Sigma_{c}^{+}(2520)}}^2:
1.8\times10^{-4}g_{{ND^{*}\Sigma_{c}^{+}(2520)}}^2:9.4\times10^{-5}g_{{ND^{*}\Sigma_{c}^{+}(2520)}}^2\;.
\end{eqnarray*}
Here $D$ exchange is most likely to dominate.

Collecting the above results, we obtain the leading contributions
to the radio of $B(\Lambda_{c}(2940)^{+}\to D^0 p)$ and
$B(\Lambda^+_c(2940)\to \Lambda^+_c \pi^+\pi^-$,
\begin{eqnarray}\label{aaaa}
&&B(\Lambda_{c}(2940)^{+}\to D^{0}p):B(\Lambda_{c}(2940)^{+}
\to\Lambda_{c}^{+}\pi^{+}\pi^{-})\\&&\nonumber=429:
(5.0\times10^{-3}g_{{N
D^{*}\Lambda_{c}^{+}(2285)}}^2+4.0g_{{ND^{*}\Sigma_{c}^{+}(2455)}}^2
+1.7\times10^{-2}g_{{ND\Sigma_{c}^{+}(2520)}}^2)\;.
\end{eqnarray}

There are no data available to determine unknown effective coupling
parameters appearing in the lagrangian eq.(\ref{lagrangian}). A
rough idea about the relative strength of these couplings can be
obtained from the use of SU(3) for the light quarks, and the heavy
quark symmetry for the heavy c-quark, since $D$, $D^*$, $\Lambda_c$
and $\Sigma_c$ all contain a relatively heavy c quark. In the heavy
quark limit the spin of heavy quark is decoupled\cite{Isgur}, so
that the dimensionless coupling constants for spin 1/2 and 3/2 heavy
baryons and spin 0 and 1 heavy mesons are approximately equal. If
applicable, one would approximately have, $g_{_{N
D^{*}\Lambda_{c}^{+}(2285)}}\simeq g_{_{N D\Sigma_{c}(2455)}}\simeq
g_{_{N D^{*}\Sigma_{c}(2455)}}\simeq g_{_{N
D\Sigma_{c}(2520)}}\simeq
g_{_{ND^{*}\Sigma_{c}(2520)}}=\mathfrak{g}$.

This approximation is, of course, very rough, but as an estimation
of the order of magnitude, they should work. Using this
approximation, we find that the contribution to
$\Lambda_{c}(2940)^{+}\to \Lambda_{c}^{+}\pi^{+}\pi^{-}$ is
dominated by the second term in eq.(\ref{aaaa}), that is dominated
by $\mathcal{M}(\Lambda_{c}^{+}(2940)\to D^{*0}p\to
\Sigma_{c}(2455)\pi\to\Lambda_{c}^{+}\pi^{+}\pi^{-})$. We then
obtain the radio of $\Lambda_{c}(2940)^{+}\to D^{0}p)$ and
$\Lambda_{c}(2940)^{+} \to\Lambda_{c}^{+}\pi^{+}\pi^{-}$ branching
radios as:
\begin{eqnarray}\label{bbbb}
&&B(\Lambda_{c}(2940)^{+}\to D^{0}p):B(\Lambda_{c}(2940)^{+}
\to\Lambda_{c}^{+}\pi^{+}\pi^{-})=429:4.0\mathfrak{g}^2.
\end{eqnarray}
One expects that the coupling $\mathfrak{g}$ involves two baryons
and a meson is similar to the couplings to $NN \pi$ and $NN\rho$
in some way. If it is comparable in size with $g_{_{NN\rho}}=3.5$,
the ratio $R=B(\Lambda_{c}(2940)^{+}
\to\Lambda_{c}^{+}\pi^{+}\pi^{-})/B(\Lambda_{c}(2940)^{+}\to
D^{0}p)$ is at order one level, and can be near to one if
$\mathfrak{g}$ is close to $g_{_{NN\pi}}=13.5$.

(2) $J^{P}(\Lambda_{c}^{+})=\frac{3}{2}^{-}$.

For the various ratios similar to the case with $J^P(\Lambda_c^+)
= 1/2^-$ discussed earlier, we have
\begin{eqnarray*}
&&|\mathcal{M}(\Lambda_{c}(2940)^{+}\to
D^{0}p)_{\pi^{0}}|^2:|\mathcal{M}(\Lambda_{c}(2940)^{+}\to
D^{0}p)_{\rho^{0}}|^2=2.0\times10^3:1,\nonumber\\
&&|\mathcal{M}(\Lambda_{c}(2940)^{+}\to
\Lambda_{c}^{+}\sigma)_{D^{*}}|^2:|\mathcal{M}(\Lambda_{c}(2940)^{+}\to
\Lambda_{c}^{+}\sigma)_{N}|^2=3.6\times10^{-2}:3.4,\nonumber\\
&&|\mathcal{M}(\Lambda_{c}(2940)^{+}\to
\Sigma_{c}(2455)\pi)_{D}|^2:|\mathcal{M}(\Lambda_{c}(2940)^{+}\to
\Sigma_{c}(2455)\pi)_{D^{*}}|^2:|\mathcal{M}(\Lambda_{c}(2940)^{+}\to
\Sigma_{c}(2455)\pi)_{N}|^2\nonumber\\
&&=4.1\times10^{-2}g_{{ND\Sigma_{c}^{+}(2455)}}^2:
6.2\times10^{-3}g_{{ND^{*}\Sigma_{c}^{+}(2455)}}^2:5.4\times10^{-2}
g_{{ND^{*}\Sigma_{c}^{+}(2455)}}^2,\nonumber\\
&&|\mathcal{M}(\Lambda_{c}(2940)^{+}\to
\Sigma_{c}(2520)\pi)_{D}|^2:|\mathcal{M}(\Lambda_{c}(2940)^{+}\to
\Sigma_{c}(2520)\pi)_{D^{*}}|^2:|\mathcal{M}(\Lambda_{c}(2940)^{+}\to
\Sigma_{c}(2520)\pi)_{N}|^2\nonumber\\
&&=8.7\times10^{-2}g_{{ND\Sigma_{c}^{+}(2520)}}^2:
9.0\times10^{-4}g_{{ND^{*}\Sigma_{c}^{+}(2520)}}^2:3.3g_{{ND^{*}\Sigma_{c}^{+}(2520)}}^2\;.
 \end{eqnarray*}

The leading contributions in the above lead to
\begin{eqnarray}\label{cccc}
&&B(\Lambda_{c}(2940)^{+}\to D^{0}p):B(\Lambda_{c}(2940)^{+}
\to\Lambda_{c}^{+}\pi^{+}\pi^{-})\nonumber\\
&&=859: (0.66g_{{N
D^{*}\Lambda_{c}^{+}(2285)}}^2+0.04g_{{ND^{*}\Sigma_{c}^{+}(2455)}}^2
+1.25g_{{ND^{*}\Sigma_{c}^{+}(2520)}}^2)\nonumber\\&&=859:1.95\mathfrak{g}^2.
\end{eqnarray}
We find that in this case the dominant contribution to
$\Lambda_c^+(2940) \to \Lambda_c^+ \pi^+\pi^-$ is from
$\Lambda_{c}(2940)^{+}\to
\Sigma_{c}(2520)\pi\to\Lambda_{c}^{+}\pi^{+}\pi^{-}$.

Although both decay modes $\Lambda^+_c(2940)\to D^0 p$ and
$\Lambda^+_c(2940)\to \Lambda^+_c \pi^+\pi^-$ have been observed
experimentally, no detailed information for
$B(\Lambda_{c}(2940)^{+}\to D^{0}p):B(\Lambda_{c}(2940)^{+}
\to\Lambda_{c}^{+}\pi^{+}\pi^{-})$ is available. But the fact that
the Belle collaboration observed $\Lambda^+_c(2940)\to \Lambda^+_c
\pi^+\pi^-$ via the intermediate $\Sigma^{++}_c(2455)$ already can
tell us interesting information about the property of
$\Lambda^+_c(2940)$. We note that with a reasonable size of
$\mathfrak{g}$ between $g_{_{NN\rho}}$ and $g_{_{NN\pi}}$, if
$\Lambda^+_c(2940)$ is a $1/2^-$ $D^{*0} p$ molecular state, there
is a sizable contribution to $\Lambda_{c}(2940)^{+}
\to\Lambda_{c}^{+}\pi^{+}\pi^{-}$ via the intermediate
$\Sigma_c(2455)$ state, but not for $3/2^-$ $D^{*0} p$ molecular
state. For a $3/2^-$ state, $\Lambda_{c}(2940)^{+}
\to\Lambda_{c}^{+}\pi^{+}\pi^{-}$ would be dominated by
$\Sigma_c(2520)$ intermediate state with a smaller branching ratio
for a similar strength for the coupling $\mathfrak{g}$. Therefore
present data support that $\Lambda^+_c(2940)$ to be a $1/2^-$ $D^*
p$ molecular state.

We now discuss some other decay properties of $\Lambda_c^+(2940)$.
If $\Lambda_{c}(2940)^{+}$ is an S-wave molecular state of
$D^{*0}p$, the binding is loose. $\Lambda_{c}(2940)^{+}$ may decay
into $D^{*0}$ and $p$ via the threshold effect. We have considered
the sequential decay of $D^{*0} \to D^0 \pi^0$ with the $\pi^0$
playing the role of a exchanged particle.  The $\pi^0$ can also
become a particle in the final state. Also
 $D^{*0}$ can decay into $D^{0}\gamma$. Thus
$\Lambda_{c}(2940)^{+}$ may have other two decay modes:
$D^{0}\pi^{0}p$ and $D^{0}\gamma p$ with sizable branching ratios
because $B(D^{*0}\to D^{0}\pi^{0})=61.9\%$ and $B(D^{*0}\to
D^{0}\gamma)=38.1\%$ are large. These decay modes should be
searched in future experiments.

If exchanged particles $\pi^{0}$ and $\rho^{0}$ in Fig.
\ref{diagram-2} are replaced by $\pi^{-}$ and $\rho^{-}$
respectively, we can get other decay modes of
$\Lambda_{c}(2940)^{+}$, i.e., $\Lambda_{c}(2940)^{+}\to
D^{+}n,\;D^{*+}n$. Since $n$ is only about 1.3 MeV heavier than $p$,
$\Lambda_c^+(2940) \to D^0 p$ and $\Lambda_c^+(2940)\to D^+ n$
should have comparable widths although the threshold suppression
factor is more sever for the latter. Search for
$\Lambda_{c}(2940)^{+}\to D^{+}n$ in future experiments should also
be carried out to understand the property of $\Lambda_c^+(2940)$.

Since $\Lambda_{c}(2940)^{+}$ in $\Lambda_{c}(2940)^{+}
\to\Lambda_{c}^{+}\pi^{+}\pi^{-}$ has been observed, the
$\Lambda_{c}(2940)^{+}$ in $\Lambda_{c}(2940)^{+}
\to\Lambda_{c}^{+}\pi^{0}\pi^{0}$ should also occur. We suggest to
look for $\Lambda_{c}(2940)^{+}$ in this channel. From our
previous discussion this decay is likely to go through
$\Lambda_c(2940)^+\to \Sigma_c(2455)^{+} \pi^{\pm} \to \Lambda_c^+
\pi^{0}\pi^{0}$ than $\Lambda_c(2940)^+\to \Sigma_c(2520)^{+}
\pi^{\pm} \to \Lambda_c^+ \pi^{0}\pi^{0}$.

In this paper we have suggested that the newly observed baryonic
state $\Lambda_{c}(2940)^{+}$ to be a $D^{*0} p$ molecular state.
The molecular structure can naturally explain why the mass is a few
MeV below the threshold, and explain the observations of
$\Lambda^+_c(2940)\to D^0 p$ and $\Lambda^+_c(2940)\to \Lambda^+_c
\pi^+\pi^-$. Observation of $\Lambda^+_c(2940)\to \Lambda^+_c
\pi^+\pi^-$ via $\Sigma_c^++(2455)$ suggests that the molecular
state is a $1/2^-$ sate. Several other decay modes of
$\Lambda_{c}(2940)^{+}$ with final products as $D^{+}n$,
$D^{0}\pi^{0}p$ , $D^{0}\gamma p$ and
$\Lambda_{c}^{+}\pi^{0}\pi^{0}$ can be used to further test the
molecular structure. We urge our experimental colleagues to carry
out such analyses.

\section*{Acknowledgements} This work is partly supported by NNSFC  and NSC.
XGH is also partially supported by NCTS.

\vspace{1cm} \noindent {\bf Appendix}\\

In this appendix we give the absorptive decay amplitude for each
diagram in  Fig. \ref{diagram-2}(\ref{diagram-3},\ref{diagram-4}).
Since $\Lambda^+_c(2940)$ is very close to the threshold of $D^{*0}
+p$, the three momenta $\vec k$ for $p$ is very small in the rest
from of $\Lambda_c(2940)$.
\\

1) $J^{P}(2940)=\frac{1}{2}^{-}$

In the following , q is the momenta of the exchange particle and
$q^2 = (k-p_3)^2=m_{N}^2+m_{3}^2-2m_{N}p_{3}^{0}$, where $m_3$ is
the mass of the particle with momentum $p_3$.

(1)$\mathcal{M}(\Lambda_{c}(2940)^{+}\to D^{*0}p\to D^{0}p)$
\begin{eqnarray*}
&&\mathcal{M}(\Lambda_{c}(2940)^{+}(p_c)\to D^{*0}(k-p_c)p(k)\to
D^{0}(p_4)p(p_3))_{\pi^{0}}( Fig. \ref{diagram-2}a)
\\=&&\bigg[\frac{|\vec{k}|}{8\pi M}\frac{g_{_{ND^{*}\Lambda_{c}^{+}(2940)}}g_{_{D^{*}D\pi}}g_{_{NN\pi}}}
{q^2-m_{\pi}^2}
(4k\cdot p_{3}-4m_{N}m_{N})\bigg]\bar{p}\Lambda_{c}^{+}\\\\
&&\mathcal{M}(\Lambda_{c}(2940)^{+}\to D^{*0}(k-p_c)p(k)\to
D^{0}(p_4)p(p_3))_{\rho^{0}}(Fig.\ref{diagram-2}b)
\\=&&\bigg[\frac{|\vec{k}|}{8\pi
M}\frac{2m_{N}m_{D^{*0}}}{M}
\frac{g_{_{ND^{*}\Lambda_{c}^{+}(2940)}}g_{_{D^{*}D\rho}}g_{_{NN\rho}}}{q^2-m_{\rho}^2}\bigg]
\bar{p}\gamma^{\mu}\gamma^{\nu}\gamma^{5}\Lambda_{c}^{+}
\epsilon_{\mu\nu\alpha\beta}p_{3}^{\alpha}p_{c}^{\beta}
\end{eqnarray*}

(2)$\mathcal{M}(\Lambda_{c}(2940)^{+}\to D^{*0}p\to
\Lambda_{c}^{+}\sigma\to\Lambda_{c}^{+}\pi^{+}\pi^{-})$
\begin{eqnarray*}
&&\mathcal{M}(\Lambda_{c}(2940)^{+}\to D^{*0}(k-p_c)p(k)\to
\Lambda_{c}^{+}(p_3)\sigma(p_4))_{D^{*}}(Fig. \ref{diagram-3}a)
\\=&&\bigg[\frac{|\vec{k}|}{8\pi M}\frac{g_{_{ND^{*}\Lambda_{c}^{+}(2940)}}
g_{_{D^{*}D^{*}\sigma}}g_{_{N D^{*}\Lambda_{c}^{+}(2285)}}}{q^2-m_{\pi}^2}
2m_{D^{*0}}(3m_{N}m_{N}-2k\cdot p_{3}+m_{N}m_{\Lambda_{c}^{+}})
\bigg]\bar{\Lambda}_{c}\gamma^{5}\Lambda_{c}^{+}\\
&&\mathcal{M}(\Lambda_{c}(2940)^{+}\to D^{*0}(k-p_c)p(k)\to
\Lambda_{c}^{+}(p_4)\sigma(p_3))_{N}(Fig.\ref{diagram-3}b)
\\=&&\bigg[\frac{|\vec{k}|}{8\pi
M}\frac{g_{_{ND^{*}\Lambda_{c}^{+}(2940)}}g_{_{NN\sigma}}g_{_{N
D^{*}\Lambda_{c}^{+}(2285)}}}{q^2-m_{\rho}^2}
(6m_{N}^2-6m_{N}M+4k\cdot
p_{4}-2m_{N}m_{\Lambda_{c}^{+}}+6m_{N}m_{N})\bigg]\bar{\Lambda}_{c}\gamma^{5}\Lambda_{c}^{+}
\end{eqnarray*}

(3)$\mathcal{M}(\Lambda_{c}^{+}(2940)\to D^{*0}p\to
\Sigma_{c}(2455)\pi\to\Lambda_{c}^{+}\pi^{+}\pi^{-})$
\begin{eqnarray*}
&&\mathcal{M}(\Lambda_{c}(2940)^{+}\to D^{*0}(k-p_c)p(k)\to
\Sigma_{c}(2455)(p_3)\pi(p_4))_{D}(Fig. \ref{diagram-4}a)
\\=&&\bigg[\frac{|\vec{k}|}{8\pi M}\frac{g_{_{ND^{*}\Lambda_{c}^{+}(2940)}}g_{_{D^{*}D\pi}}g_{_{N\Sigma_{c}(2455)
D}}}{q^2-m_{D}^2} (4k\cdot
p_{3}-4m_{N}m_{\Sigma_{c}})\bigg]\bar{\Sigma}_{c}\Lambda_{c}^{+}\\\\
&&\mathcal{M}(\Lambda_{c}(2940)^{+}\to
D^{*0}(k-p_c)p(k)\to\Sigma_{c}(2455)(p_3)\pi(p_4))_{D^{*}}(Fig.\ref{diagram-4}b)
\\=&&\bigg[\frac{|\vec{k}|}{8\pi
M}\frac{2m_{N}m_{D^{*0}}}{M}
\frac{g_{_{ND^{*}\Lambda_{c}^{+}(2940)}}g_{_{D^{*}D^{*}\pi}}g_{_{N\Sigma_{c}(2455)
D^{*}}}}{q^2-m_{D^{*}}^2}\bigg]\bar{\Sigma}_{c}
\gamma^{\mu}\gamma^{\nu}\Lambda_{c}^{+}
\epsilon_{\mu\nu\alpha\beta}p_{3}^{\alpha}p_{c}^{\beta}\\\\
&&\mathcal{M}(\Lambda_{c}(2940)^{+}\to D^{*0}(k-p_c)p(k)\to
\Sigma_{c}(2455)(p_4)\pi(p_3))_{N}(Fig. \ref{diagram-4}c)
\\=&&\bigg[\frac{|\vec{k}|}{8\pi M}\frac{g_{_{ND^{*}\Lambda_{c}^{+}(2940)}}g_{_{NN\pi}}g_{_{N\Sigma_{c}(2455)
D^{*}}}}{q^2-m_{N}^2}(6m_{N}^2-6m_{N}M+4k\cdot
p_{4}+2m_{N}m_{\Sigma_{c}^{+}}-6m_{N}m_{N})\bigg]\bar{\Sigma}_{c}\Lambda_{c}^{+},\\\\
\end{eqnarray*}

(4)$\mathcal{M}(\Lambda_{c}(2940)^{+}\to D^{*0}p\to
\Sigma_{c}(2520)\pi\to\Lambda_{c}^{+}\pi^{+}\pi^{-})$
\begin{eqnarray*}
&&\mathcal{M}(\Lambda_{c}(2940)^{+}\to D^{*0}(k-p_c)p(k)\to
\Sigma_{c}(2520)(p_3)\pi(p_4))_{D}(Fig. \ref{diagram-4}a)
\\=&&\bigg[\frac{|\vec{k}|}{8\pi M m_{D}}\frac{g_{_{ND^{*}\Lambda_{c}^{+}(2940)}}g_{_{D^{*}D\pi}}g_{_{N\Sigma_{c}(2520)
D}}}{q^2-m_{D}^2} (4k\cdot
p_{3}+4m_{N}m_{\Sigma_{c}})\bigg]\bar{\Sigma}_{c}^{\mu}k_{\mu}\gamma^{5}\Lambda_{c}^{+}\\
&&\mathcal{M}(\Lambda_{c}(2940)^{+}\to
D^{*0}(k-p_c)p(k)\to\Sigma_{c}(2520)(p_3)\pi(p_4))_{D^{*}}(Fig.\ref{diagram-4}b)
\\=&&\bigg[\frac{|\vec{k}|}{8\pi M}\frac{2m_{N}m_{D^{*0}}}{M}
\frac{g_{_{ND^{*}\Lambda_{c}^{+}(2940)}}g_{_{D^{*}D^{*}\pi}}g_{_{N\Sigma_{c}(2520)
D^{*}}}}{q^2-m_{D^{*}}^2}\bigg]\bar{\Sigma}_{c}^{\mu}\gamma^{\nu}\Lambda_{c}^{+}\epsilon_{\mu\nu\alpha\beta}p_{3}^{\alpha}p_{c}^{\beta}\\\\
&&\mathcal{M}(\Lambda_{c}(2940)^{+}\to D^{*0}(k-p_c)p(k)\to
\Sigma_{c}(2520)(p_4)\pi(p_3))_{N}(Fig. \ref{diagram-4}c)
\\=&&\bigg[\frac{|\vec{k}|}{8\pi M}\frac{g_{_{ND^{*}\Lambda_{c}^{+}(2940)}}g_{_{NN\pi}}g_{_{N\Sigma_{c}(2520)
D^{*}}}}{q^2-m_{N}^2}(6m_{N}^2-6m_{N}M+4k\cdot
p_{4}+2m_{N}m_{\Sigma_{c}^{+}}-6m_{N}m_{N})\bigg]\bar{\Sigma}_{c}^{\mu}k_{\mu}\gamma^{5}\Lambda_{c}^{+}\\\\
\end{eqnarray*}

2) $J^{P}(2940)=\frac{3}{2}^{-}$

(1)$\mathcal{M}(\Lambda_{c}(2940)^{+}\to D^{*0}p\to D^{0}p)$
\begin{eqnarray*}
&&\mathcal{M}(\Lambda_{c}(2940)^{+}\to D^{*0}(k-p_c)p(k)\to
D^{0}(p_3)p(p_4))_{\pi^{0}}(Fig. \ref{diagram-2}a)
\\=&&\bigg[\frac{|\vec{k}|}{8\pi M}
\frac{g_{_{ND^{*}\Lambda_{c}^{+}(2940)}}g_{_{D^{*}D\pi}}g_{_{NN\pi}}}{q^2-m_{\pi}^2}
4m_{N}\bigg]
\bar{P}\gamma^{5}\Lambda_{c}^{+\mu}{p_{3\mu}}\\\\
&&\mathcal{M}(\Lambda_{c}(2940)^{+}\to D^{*0}(k-p_c)p(k)\to
D^{0}(p_3)p(p_4))_{\rho^{0}}(Fig.\ref{diagram-2}b)
\\=&&\bigg[\frac{|\vec{k}|}{8\pi
M}\frac{2m_{N}m_{D^{*0}}}{M}
\frac{g_{_{ND^{*}\Lambda_{c}^{+}(2940)}}g_{_{D^{*}D\rho}}g_{_{NN\rho}}}{q^2-m_{\rho}^2}\bigg]
\bar{P}\gamma^{\mu}\Lambda_{c}^{+\nu}
\epsilon_{\mu\nu\alpha\beta}p_{3}^{\alpha}p_{c}^{\beta}
\end{eqnarray*}

(2)$\mathcal{M}(\Lambda_{c}(2940)^{+}\to D^{*0}p\to
\Lambda_{c}^{+}\sigma\to\Lambda_{c}^{+}\pi^{+}\pi^{-})$
\begin{eqnarray*}
&&\mathcal{M}(\Lambda_{c}(2940)^{+}\to D^{*0}(k-p_c)p(k)\to
\Lambda_{c}^{+}(p_3)\sigma(p_4))_{D^{*}}(Fig. \ref{diagram-3}a)
\\=&&\bigg[\frac{|\vec{k}|}{8\pi M}
\frac{g_{_{ND^{*}\Lambda_{c}^{+}(2940)}}g_{_{D^{*}D^{*}\sigma}}g_{_{N
D^{*}\Lambda_{c}^{+}(2285)}}}{q^2-m_{D^{*}}^2}2m_{N}m_{D^{*0}} \bigg]\bar{\Lambda}_{c}\Lambda_{c}^{+\mu}{p_{3\mu}}\\
&&\mathcal{M}(\Lambda_{c}(2940)^{+}\to D^{*0}(k-p_c)p(k)\to
\Lambda_{c}^{+}(p_4)\sigma(p_3))_{N}(Fig.\ref{diagram-3}b)
\\=&&\bigg[\frac{|\vec{k}|}{8\pi
M}\frac{g_{_{ND^{*}\Lambda_{c}^{+}(2940)}}g_{_{NN\sigma}}g_{_{N
D^{*}\Lambda_{c}^{+}(2285)}}}{q^2-m_{N}^2} 4m_{N}\bigg]
\bar{\Lambda}_{c}\Lambda_{c}^{+\mu}{p_{4\mu}}
\end{eqnarray*}

(3)$\mathcal{M}(\Lambda_{c}^{+}(2940)\to D^{*0}p\to
\Sigma_{c}(2455)\pi\to\Lambda_{c}^{+}\pi^{+}\pi^{-})$
\begin{eqnarray*}
&&\mathcal{M}(\Lambda_{c}(2940)^{+}\to D^{*0}(k-p_c)p(k)\to
\Sigma_{c}(2455)(p_3)\pi(p_4))_{D}(Fig. \ref{diagram-4}a)
\\=&&\bigg[\frac{|\vec{k}|}{8\pi M}\frac{g_{_{ND^{*}\Lambda_{c}^{+}(2940)}}g_{_{D^{*}D\pi}}g_{_{N\Sigma_{c}(2455)
D}}}{q^2-m_{D}^2} 4m_{N}\bigg]\bar{\Sigma}_{c}\gamma^{5}\Lambda_{c}^{+\mu}{p_{3\mu}}\\\\
&&\mathcal{M}(\Lambda_{c}(2940)^{+}\to
D^{*0}(k-p_c)p(k)\to\Sigma_{c}(2455)(p_3)\pi(p_4))_{D^{*}}(Fig.\ref{diagram-4}b)
\\=&&\bigg[\frac{|\vec{k}|}{8\pi
M}\frac{2m_{N}m_{D^{*0}}}{M}
\frac{g_{_{ND^{*}\Lambda_{c}^{+}(2940)}}g_{_{D^{*}D^{*}\pi}}g_{_{N\Sigma_{c}(2455)
D^{*}}}}{q^2-m_{D^{*}}^2}\bigg]\bar{\Sigma}_{c}\gamma^{\mu}\Lambda_{c}^{+\nu}
\epsilon_{\mu\nu\alpha\beta}p_{3}^{\alpha}p_{c}^{\beta}\\\\
&&\mathcal{M}(\Lambda_{c}(2940)^{+}\to D^{*0}(k-p_c)p(k)\to
\Sigma_{c}(2455)(p_4)\pi(p_3))_{N}(Fig. \ref{diagram-4}c)
\\=&&\bigg[\frac{|\vec{k}|}{8\pi M}\frac{g_{_{ND^{*}\Lambda_{c}^{+}(2940)}}g_{_{NN\pi}}g_{_{N\Sigma_{c}(2455)
D^{*}}}}{q^2-m_{N}^2}4m_{N}\bigg]\bar{\Sigma}_{c}\gamma^{5}\Lambda_{c}^{+\mu}{p_{3\mu}}\\\\
\end{eqnarray*}

(4)$\mathcal{M}(\Lambda_{c}(2940)^{+}\to D^{*0}p\to
\Sigma_{c}(2520)\pi\to\Lambda_{c}^{+}\pi^{+}\pi^{-})$
\begin{eqnarray*}
&&\mathcal{M}(\Lambda_{c}(2940)^{+}\to D^{*0}(k-p_c)p(k)\to
\Sigma_{c}(2520)(p_3)\pi(p_4))_{D}(Fig. \ref{diagram-4}a)
\\=&&\bigg[\frac{|\vec{k}|}{8\pi M m_{D}}\frac{g_{_{ND^{*}\Lambda_{c}^{+}(2940)}}g_{_{D^{*}D\pi}}g_{_{N\Sigma_{c}(2520)
D}}}{q^2-m_{D}^2}4m_{N}\bigg]\bar{\Sigma}_{c}^{\mu}k_{\mu}\Lambda_{c}^{+\nu}{p_{3\nu}}\\\\
&&\mathcal{M}(\Lambda_{c}(2940)^{+}\to
D^{*0}(k-p_c)p(k)\to\Sigma_{c}(2520)(p_3)\pi(p_4))_{D^{*}}(Fig.\ref{diagram-4}b)
\\=&&\bigg[\frac{|\vec{k}|}{8\pi
M}\frac{2m_{N}m_{D^{*0}}}{M}
\frac{g_{_{ND^{*}\Lambda_{c}^{+}(2940)}}g_{_{D^{*}D^{*}\pi}}g_{_{N\Sigma_{c}(2520)
D^{*}}}}{q^2-m_{D^{*}}^2}\bigg]\bar{\Sigma}_{c}^{\mu}\gamma^{5}\Lambda_{c}^{+\nu}
\epsilon_{\mu\nu\alpha\beta}p_{3}^{\alpha}p_{c}^{\beta}\\\\
&&\mathcal{M}(\Lambda_{c}(2940)^{+}\to D^{*0}(k-p_c)p(k)\to
\Sigma_{c}(2520)(p_4)\pi(p_3))_{N}(Fig. \ref{diagram-4}c)
\\=&&\bigg[\frac{|\vec{k}|}{8\pi M}\frac{g_{_{ND^{*}\Lambda_{c}^{+}(2940)}}g_{_{NN\pi}}g_{_{N\Sigma_{c}(2520)
D^{*0}}}}{q^2-m_{N}^2}2m_{N}(M-m_{\Sigma_{c}^{+}})\bigg]\bar{\Sigma}_{c}^{\mu}{\Lambda_{c}^{+}}_{\mu},\\
\end{eqnarray*}


\begin{thebibliography}{99}
\bibitem{babar}B. Aubert et al., Babar Collaboration, Phys. Rev. Lett. {\bf 98}, 012001 (2007).


\bibitem{CLEO} M. Artuso et al., CLEO Collaboration, Phys. Rev.
Lett. {\bf 86}, 4479 (2001).

\bibitem{belle} K. Abe et al., the Belle collaboration, arXiv:
hep-ex/0608043.

\bibitem{2880-1}S. Migura, D. Merten, B. Metsch and H.R. Petry, Eur. Phys. J. {\bf A 28}, 41 (2006).

\bibitem{2880-2}D. Pirjol and T.M. Yan, Phys. Rev. {\bf D 56}, 5483
(1997).

\bibitem{2880-3} A.E. Blechman, A.F. Falk, D. Pirjol and J.M. Yelton, Phys.
Rev. {\bf D 67}, 074033 (2003).

\bibitem{Tuan} S.F. Tuan, Phys. Rev. {\bf D15} (1977) 3478; J.
Rosner and S.F. Tuan, Phys. Rev. {\bf D27} (1983) 1544; S.F. Tuan,
Phys. Lett. {\bf B473} (2000) 136.

\bibitem{980-1}J. Weinstein and N. Isgur, Phys. Rev. Lett. {\bf 48},
659 (1982).

\bibitem{980-2}J. Weinstein and N. Isgur, Phys. Rev. {\bf D 27}, 588
(1983).

\bibitem{980-3}J. Weinstein and N. Isgur, Phys. Rev. {\bf D 41}, 2236
(1990).

\bibitem{RGG}A.D. Rujula, H. Georgi and S.L. Glashow, Phys. Rev.
Lett. {\bf 38}, 317 (1977).

\bibitem{Y4260} X. Liu, X.Q. Zeng and X.Q. Li, Phys. Rev. {\bf
D 72}, 054023 (2005).

\bibitem{Yuan} C.Z. Yuan, P. Wang and X.H. MO, Phys. Lett. {\bf
B 634}, 399 (2006).

\bibitem{Okun} M.B. Voloshin and L.B. Okun, JETP Lett. {\bf 23},
333 (1976).

\bibitem{Liu} X.G. He, X.Q. Li, X. Liu and X.Q. Zeng, Eur. Phys. J. {\bf C 44}, 419
(2005).

\bibitem{Landau} V. Berestetskii, E. Lifshitz and L. Pitaevskii,
{\it Quantum Electrodynamics}, Pergamon Press, 1982, New York.


\bibitem{formfactor}M.P. Locher, Y. Lu and B.S. Zou, Z. Phys. {\bf
A 347}, 281 (1994); X.Q. Li, D.V. Bugg and B.S. Zou, Phys. Rev. {\bf
D 55}, 1423 (1997).

\bibitem{footnote} In the literature, different types of the form factors are
offered, and we have tried several of them and find that by
slightly adjusting the corresponding phenomenological parameters,
the results obtained with different types are close.

\bibitem{cheng}H.Y. Cheng, C.K. Chua and A. Soni, Phys. Rev. {\bf D71}, 014030 (2005).

\bibitem{chung}S.U. Chung, Phys. Rev. {\bf D 48}, 1225 (1993).

\bibitem{PDG}S. Eidelman et al., Phys. Lett. {\bf B 592}, 1 (2004).

\bibitem{Isgur}N. Isgur and M. Wise, Phys. Rev. Lett. {\bf 66}, 1130
(1991); N. Isgur and M. Wise, Phys. Lett. {\bf B 232} 113 (1989);
{\bf B 237} 527 (1990).


\end{thebibliography}
\end{document}